\documentclass{iopconfser}
\usepackage{graphicx}
\usepackage{subcaption}
\usepackage{amsmath}
\usepackage{placeins}
\usepackage{physics}
\usepackage{orcidlink}
\usepackage{etoolbox}

\begin{document}

\title{Light deflection in the gravimagnetic dipole spacetime}

\author{Cl\'ementine Dassy, Jan Govaerts}

\affil{Centre for Cosmology, Particle Physics and Phenomenology (CP3), UCLouvain, Louvain-la-Neuve, Belgium}


\email{clementine.dassy@uclouvain.be}
\myorcid{CD: \orcidlinkc{0000-0002-0965-7848}, JG: \orcidlinkc{0000-0002-8430-5180}}

\begin{abstract}
The gravimagnetic dipole is an asymptotically flat, stationary, axisymmetric vacuum solution to Einstein's General Relativity describing two non-extreme black holes with equal masses and opposite NUT charges connected by a Misner string. The string’s tension can be set to zero by choosing the black hole separation accordingly, yielding a stable system of oppositely rotating black holes at fixed distance. Numerical simulations of massless particle geodesics reveal gravitational lensing effects for extended sources at infinity on the equatorial plane or on the vertical axis.
\end{abstract}
\section{Introduction}
The gravimagnetic dipole spacetime is a stationary axisymmetric solution to Einstein's vacuum equations\cite{ClementG}. In Weyl-Papapetrou coordinates $(ct, \rho, \phi, z) = (x^0, x^i)$, the metric element is given by
\begin{equation}
\dd s^2 = -f(c \dd t - \omega \dd \phi)^2 + f^{-1} \left( e^{2 \gamma} (\dd \rho^2 + \dd z^2) + \rho^2 \dd \phi^2 \right)
\end{equation}
with $f , \omega$ and $e^{2 \gamma} $ depending only on $(\rho, z)$.
This geometry involves three parameters: mass $m$, NUT parameter $\nu$ and separation $2k$, as shown in Fig.~1(a). It describes two rotating black holes if the horizons extremities $\alpha_\pm = \sqrt{m^2 + k^2 - \nu^2 \pm 2 d}$ on the $z$ axis are real and positive, where $d = \sqrt{m^2 k^2 + \nu^2 (k^2 - m^2)}$; equilibrium requires a tensionless Misner string, fixing 
\begin{equation}
k(\nu, m) = \sqrt{\frac{m^6 + 3 m^4 \nu^2 + m^2 \nu \sqrt{4 m^6 + 9 m^4 \nu^2 + 2 m^2 \nu^4 + \nu^6}}{m^4 - \nu^4}}.
\end{equation}
All quantities are adimensional with $Gmc^2 = 1 = m$.

We determine the photon deflection angles for different values of $\nu$, first in the equatorial plane in Section \ref{section:equatorial}, then for photons incoming along the positive $z$-axis in Section \ref{section:axial}.
\begin{figure}[h]
\centering
\includegraphics[width=1\textwidth]{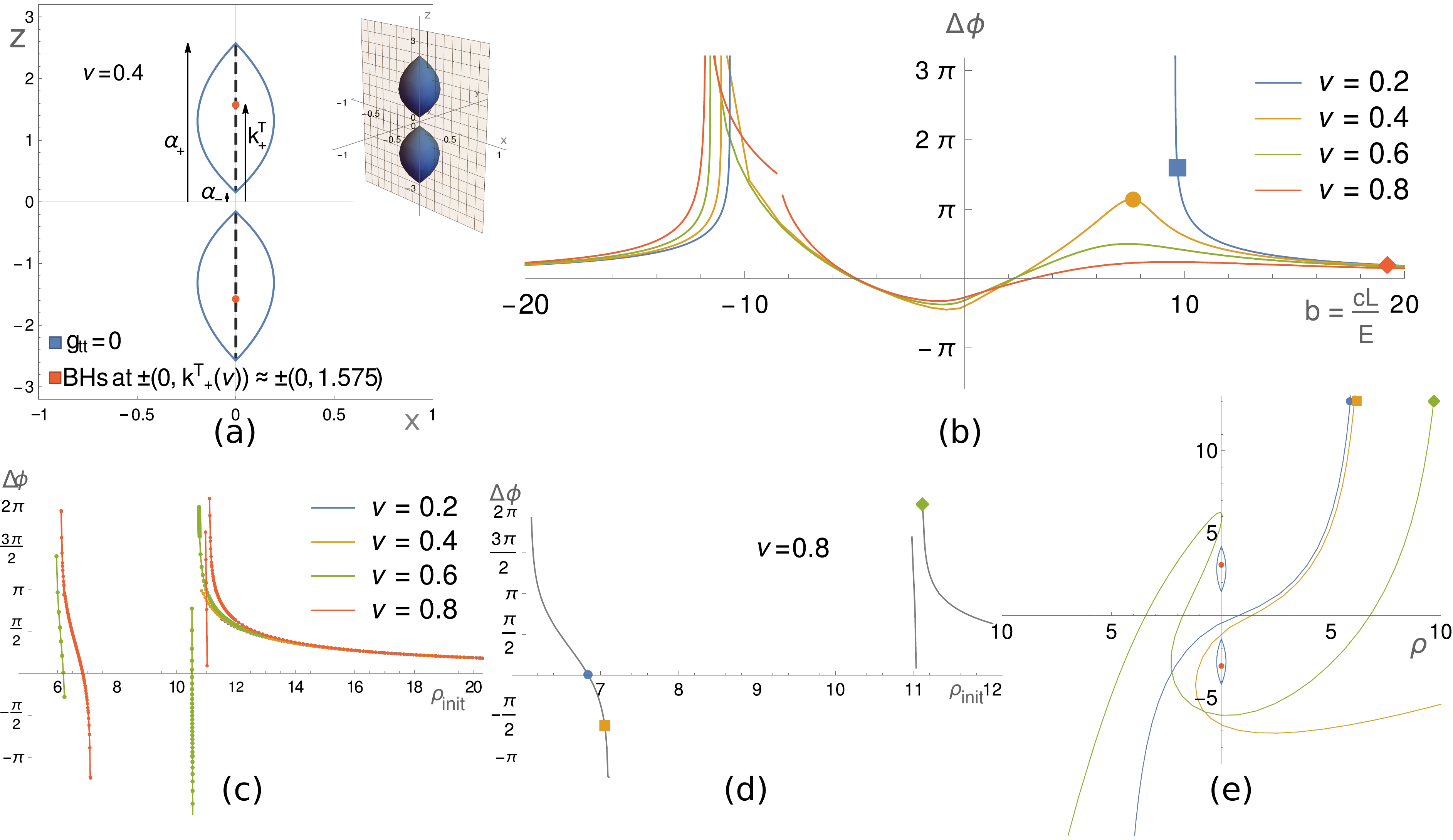}
\caption{\textbf{(a)} Slice of the $XZ$ plane showing the black holes (orange dots), their horizons (dashed lines) and static limits (blue lines)\cite{ClementG,Misner} for $\nu = 0.4$ and $k \approx 1.575$. The real, positive $\alpha_\pm$ mark the horizon extremities on the $z$ axis. \textbf{(b)} Deflection angle as functions of $b$ for four different values of $\nu$. The singularity at $\rho =0$ appears as a discontinuity in $\Delta \phi$ near $b \approx -10$. For $\left\vert b\right\vert \gg k$, the curves resemble those for a single rotating black hole. Vertical asymptotes correspond to circular orbits. The zeros in $\Delta \phi$ correspond to photons that can pass between the black holes with their trajectory unaffected. \textbf{(c)} Deflection angle as a function of $\rho_{\mbox{init}}$ for four values of $\nu$. The rightmost part resembles single rotating black holes, the leftmost part shows that for $\nu$ large enough, photons can pass in between the black holes. \textbf{(d)} Zoom for $\nu = 0.8$. The zeros of $\Delta \phi$ indicate that photons can pass between the black holes and continue mostly unaffected. The markers refer to the trajectories depicted in (e). \textbf{(e)} Three example trajectories from (d). The blue circle and the orange square pass between the black holes then go to infinity, while the green diamond rotates in the $XY$ plane while curving around the upper black hole. \label{fig}}
\end{figure}

\section{Light deflection in the equatorial plane}\label{section:equatorial}
In the equatorial plane, the geodesic equations and the constraint ($p_\mu  = g_{\mu \nu} \dot{x}^\nu, \  g^{\mu \nu} p_\mu p_\nu = 0, \ p_z = 0 = z$) yield as conserved quantities the energy $p_t = - E/c$ and the angular momentum $p_\phi = L$,  with the overdot denoting the derivative relative to the worldline parameter $u$, $\dot{x}^\mu = \dv*{x^\mu}{u}$. One can define the deflection angle $\Delta \phi$ that depends on the effective potential $V(\rho)$\cite{IgataT}:
\begin{equation}
\Delta \phi =  2 \int^\infty_{\rho_0} \frac{\left\vert \dd \rho \right\vert}{\sqrt{-V(\rho)} }- \pi, \qquad \left(\dv{\rho}{\phi}\right)^{2}  + V(\rho) = 0,\quad V(\rho) = \frac{\rho^2}{e^{2 \gamma}} \left( 1 - \frac{\rho^2}{f^2} \frac{1}{(b-\omega)^2}\right) 
\end{equation}
and on the closest approach radius $\rho_0$. The impact parameter $b$ verifies $V(\rho) = 0$, so that $b^\pm = cL/E = \omega(\rho_0) \pm \rho_0/f(\rho_0)$,
with the choice of sign corresponding to prograde $(+)$ and retrograde $(-)$ photons.

This integral is numerically computed for a range of values of $\rho_0$ with its results plotted in Fig.~\ref{fig}(b) as four functions of $b$ for four different values of $\nu$. Note that the singularity in $\rho =0$ is shown as a discontinuity in $\Delta \phi$ around $b \approx -10$ for each curve. For $\left\vert b\right\vert \gg k$, the deflection curves behave as would \pagebreak be expected in the case of a single rotating black hole. The vertical asymptotes correspond to circular orbits for the photons. The zeros in $\Delta \phi$ correspond to photons that can pass between the black holes with their trajectory unaffected.

\section{Light deflection along the vertical axis}\label{section:axial}

Looking at the deflection for photons incoming along the $z$-axis, the same method cannot be used anymore, since the deviation will mostly impact the radial position. To still get an idea of what happens, photons are generated with a panel of initial conditions and their deflection $\Delta \phi$, defined\cite{MendozaSantibanez} as
\begin{equation}
\cos \Delta \phi = \frac{\dot{x}_i^k \dot{x}_{f, k}}{\left \vert \dot{x}_i\right\vert \left\vert \dot{x}_f\right\vert}, \mbox{ with } \dot{x}^k = \dv{x^k}{x^0}, \left\vert \dot{x} \right\vert = \sqrt{\dot{x}^k \dot{x}_k}, \dot{x}_{i,f} \mbox{ respectively the initial and final velocities,}
\end{equation}
is plotted on Fig.~\ref{fig}(c) as a function of the initial radius $\rho_{\mbox{init}}$. Let us note that the rightmost part of the graph resembles the one of a single rotating black hole, with vertical asymptotes appearing around $\rho_{\mbox{init}} = 11$. However, the leftmost part, shown on a bigger scale in Fig.~\ref{fig}(d), indicates that photons can pass between the black holes and curl back around one or the other. The three markers (respectively blue dot, orange square and green diamond) correspond to the trajectories shown in Fig.~\ref{fig}(e), marking the earliest position shown.

\section{Conclusion}
Having looked at these two cases of light deﬂection, we can conclude that interesting and intriguing patterns do appear for a distant observer whenever there is a gravimagnetic dipole system in the line of sight.
\FloatBarrier

\providecommand{\newblock}{}


\begin{thebibliography}{1}
\expandafter\ifx\csname url\endcsname\relax
  \def\url#1{{\tt #1}}\fi
\expandafter\ifx\csname urlprefix\endcsname\relax\def\urlprefix{URL }\fi
\providecommand{\eprint}[2][]{\url{#2}}

\bibitem{ClementG}
Clément G 2021 {\em Class. Quantum Gravity\/} {\bf 38} 075003
  \urlprefix\url{https://doi.org/10.1088/1361-6382/abe4ed}

\bibitem{Misner}
Misner C~W, Thorne K~S and Wheeler J~A 1973 {\em {Gravitation}\/} (W. H.
  Freeman, Princeton University Press)

\bibitem{IgataT}
Igata T 2025 Deflection angle in the strong deflection limit and quasinormal
  modes in stationary axisymmetric spacetimes (\textit{Preprint}
  \eprint{2505.01848}) \urlprefix\url{https://arxiv.org/abs/2505.01848}

\bibitem{MendozaSantibanez}
Mendoza S and Santiba\~{n}ez Armenta M 2024 {\em Int. J. Geom. Methods Mod.
  Phys.\/} {\bf 21} 2450209
  \urlprefix\url{https://doi.org/10.1142/S0219887824502098}

\end{thebibliography}
\end{document}